\documentclass[]{aa}
\pdfoutput=1
\usepackage[varg]{txfonts}
\usepackage{natbib}
\usepackage{txfonts}
\usepackage[labelfont=bf]{caption}
\usepackage{adjustbox}
\usepackage[super]{nth}
\usepackage[]{siunitx}

\begin{document}
\urlstyle{same}
\title{The spectral energy distribution of the
 redshift 7.1 quasar ULAS J1120+0641} 
\author{
R. Barnett\inst{\ref{inst1}}
\and S. J. Warren\inst{\ref{inst1}}
\and M. Banerji\inst{\ref{inst2},\ref{inst3},\ref{inst4}}
\and R. G. McMahon\inst{\ref{inst3},\ref{inst4}}
\and P. C. Hewett\inst{\ref{inst3}}
\and D. J. Mortlock\inst{\ref{inst1},\ref{inst5}}
\and C. Simpson\inst{\ref{inst6}}
\and B. P. Venemans\inst{\ref{inst7}}
\and K. Ota\inst{\ref{inst4},\ref{inst8}}
\and T. Shibuya\inst{\ref{inst9}}
}
\institute{Astrophysics Group, Blackett Laboratory, Imperial College
  London, Prince Consort Road, London SW7 2AZ, UK. \label{inst1} 
\and Department of Physics \& Astronomy, University College London,
Gower Street, London WC1E 6BT, UK\label{inst2} 
\and Institute of Astronomy, University of Cambridge, Madingley Road,
Cambridge CB3 0HA, UK\label{inst3} 
\and Kavli Institute for Cosmology, University of Cambridge, Madingley
Road, Cambridge CB3 0HA, UK\label{inst4} 
\and Department of Mathematics, Imperial College London, London SW7 2AZ, UK. \label{inst5} 
\and Astrophysics Research Institute, Liverpool John Moores University, Liverpool Science Park, 146 Brownlow Hill, Liverpool L3 5RF, UK\label{inst6} 
\and Max Planck Institut f{\"u}r Astronomie, K{\"o}nigstuhl 17,
D-69117 Heidelberg, Germany \label{inst7} 
\and Cavendish Laboratory, University of Cambridge, 19 J.J. Thomson Avenue, Cambridge CB3 0HE, UK \label{inst8} 
\and Institute for Cosmic Ray Research, The University of Tokyo, 5-1-5 Kashiwanoha, Kashiwa, Chiba 277-8582, Japan \label{inst9}
}
\date{Received <> / Accepted <>}

\abstract{We present new observations of the highest-redshift quasar
  known, ULAS J1120+0641, redshift $z=7.084$, obtained in the optical,
  at near-, mid-, and far-infrared wavelengths, and in the sub-mm. We
  combine these results with published X-ray and radio observations to
  create the multiwavelength spectral energy distribution (SED), with the goals of measuring the
  bolometric luminosity $L_{\rm bol}$, and quantifying the respective
  contributions from the AGN and star formation. We find three
  components are needed to fit the data over the wavelength range
  $0.12-1000\,\mu$m: the unobscured quasar accretion disk and broad-line
  region, a dusty clumpy AGN torus, and a cool 47K modified black body
  to characterise star formation. Despite the low signal-to-noise
  ratio of the new long-wavelength data, the normalisation of any
  dusty torus model is constrained within $\pm40\%$. We measure a
  bolometric luminosity $L_{\rm
    bol}=2.6\pm0.6\times10^{47}$\,erg\,s$^{-1}=6.7 \pm
  1.6\times10^{13}L_{\odot}$, to which the three components contribute
  $31\%,32\%,3\%$, respectively, with the remainder provided by the
  extreme UV $<0.12\,\mu$m. We tabulate the best-fit model SED. We use
  local scaling relations to estimate a star formation rate (SFR) in
  the range $60-270\,{\rm M}_\odot$/yr from the [C\,\textsc{ii}] line
  luminosity and the $158\,\mu$m continuum luminosity. An analysis of
  the equivalent widths of the [C\,\textsc{ii}] line in a sample of
  $z>5.7$ quasars suggests that these indicators are
  promising tools for estimating the SFR in high-redshift quasars in
  general. At the time observed the black hole was growing in mass
  more than 100 times faster than the stellar bulge, relative to the
  mass ratio measured in the local universe, i.e. compared to
  ${M_{\rm BH}}/{M_{\rm bulge}} \simeq 1.4\times10^{-3}$, for ULAS
  J1120+0641 we measure ${\dot{M}_{\rm BH}}/{\dot{M}_{\rm bulge}}
  \simeq 0.2$.}


\maketitle

\section{Introduction}
\label{sec:intro}

\begin{figure*}[t]
\centering
\advance\leftskip-3cm
\advance\rightskip-3cm
\includegraphics[scale = 0.27, trim = 0cm 0.0cm 0cm 0.0cm, clip]{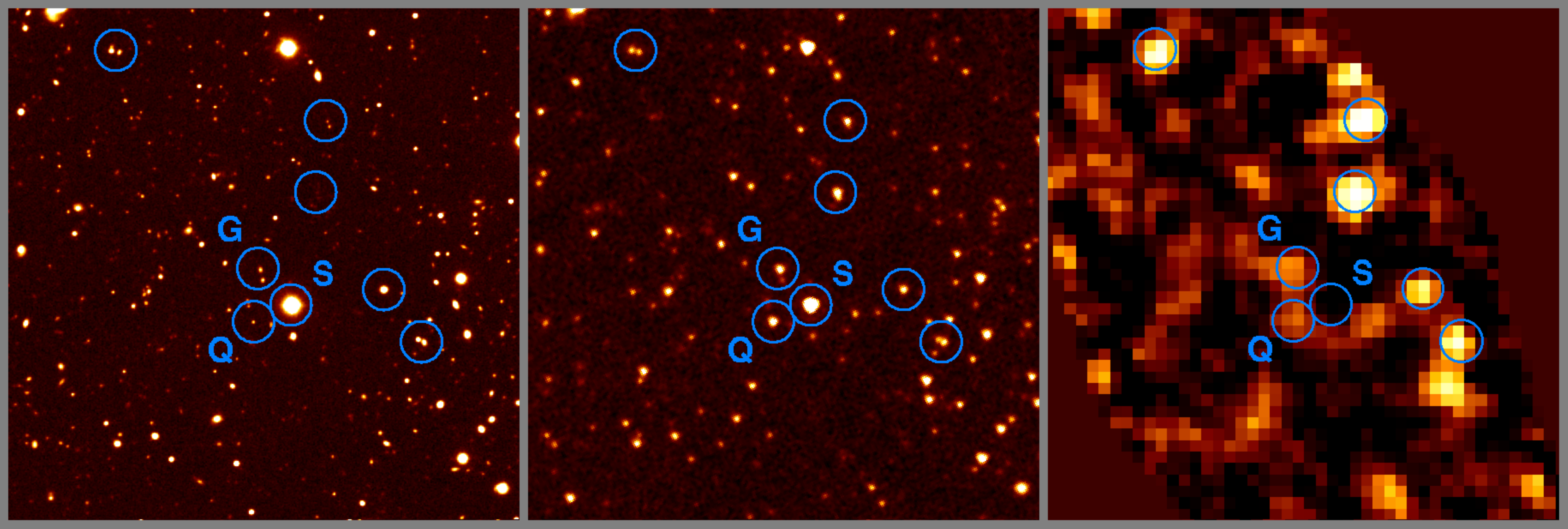}
\caption{Images of the quasar field. Left: Subaru $z^{\prime}$ image,
  Middle: \emph{Spitzer} Ch2, Right: \emph{Herschel} PACS 100 and
  160\,$\mu$m images combined (pixel scale 4$\arcsec$). The field of
  view is $3\times3$ arcmin. N is up and E to the left. The five
  unlabelled circles are examples of sources detected in PACS. Q marks
  the quasar, S the nearby bright star (\emph{Spitzer} photometry,
  \S\ref{sec:spitzer}), and G the nearby galaxy (PACS photometry,
  \S\ref{sec:pacs}).}
\label{stamp}
\end{figure*} 

The most distant known quasars, seen at redshifts of $z>6$
\citep[e.g.][]{Fan2001, Fan2004, Jiang2009, Willott2010, Mortlock2011,
  Venemans2013}, are the brightest non-transient sources at these
early times, and so are valuable for measuring the conditions in the
inter-galactic medium in the first billion years after the Big Bang,
and have been used to chart the progress of cosmic reionisation
\citep{Fan2006, Bolton2011}. These sources are also interesting in
themselves, because of the short time, a few hundred Myr, available to
grow the nuclear supermassive black holes, and to enrich the
broad-line region to super-solar metallicities. The discovery of black
holes of mass \(>10^9\)\,M\(_\odot\) at $z>6$
\citep{Willott2003}, and now $z>7$ \citep{Mortlock2011}, poses a
challenge to the standard model of their formation by
Eddington-limited growth from stellar-mass seed black holes
\citep[e.g.][]{Volonteri2010}, leading several authors to investigate
the formation of massive black-hole seeds M \(>10^4\) M\(_\odot\)
through direct collapse \citep{Loeb1994, Begelman2006, Regan2009}. In
a similar vein, the lack of evolution in the metallicity of quasars at
any redshift out to $z\simeq7$, their supersolar metallicities, and the
constancy of the ratio of Fe to $\alpha$ elements, imply a high rate
of star formation at much higher redshift and provide constraints on
the form of the IMF at these early times \citep{Dietrich2003a,
  Dietrich2003b, Venkatesan2004, deRosa2014}.

There is a close similarity between the cosmic histories of black hole
accretion and star formation. The integrated luminosity in quasars and
the universal star formation rate in galaxies both increase strongly
from today back to $z\sim2-3$ \citep{Croom2004, Hopkins2006}, and
decline at higher redshifts \citep{Fan2001, McGreer2013, Bouwens2011},
and there is a correlation between the mass of the central black hole
and the galaxy bulge mass \citep{Magorrian1998, Ferrarese2000,
  Haring2004}. Measuring the properties of the highest-redshift
quasars can provide clues to the mechanisms responsible for the origin
of this relation \citep[e.g.][]{Kauffmann2000}, and to the black hole
seeding mechanism \citep[e.g.][]{Natarajan2014}.

New observing facilities, especially the \emph{Herschel} Space
Observatory \citep{Pilbratt2010}, and the Submillimetre Common-User
Bolometer Array 2 \citep[SCUBA-2;][]{Holland2013}, have made it
possible to obtain photometry of high-redshift sources at far-infrared
and sub-mm wavelengths. The recent compilation by \citet{Leipski2014}
of \emph{Spitzer} and \emph{Herschel} observations of 69 $z>5$ quasars
is a landmark in the study of the spectral energy distributions (SEDs)
of the highest-redshift quasars. To analyse the SEDs they perform
multi-component fits, including a clumpy torus model. They found that
modelling the $\sim15\%$ of sources detected with \emph{Herschel} at
$250-500\mu$m requires an additional $10^{13}L_\odot$ cold $\sim50$K
component that is likely attributable to star formation.

In this paper we present the multiwavelength (X-ray to radio) SED of
the highest redshift quasar known, ULAS J1120+0641, $z=7.084$
\citep{Mortlock2011, Venemans2012}. Previously published observations
of this source include X-ray data acquired with \emph{Chandra} and \emph{XMM-Newton}
(\citealt{Page2014}; see also \citealt{Moretti2014}), ground-based and
\emph{Hubble Space Telescope} (\emph{HST}) optical and near-infrared imaging
\citep{Mortlock2011, Simpson2014}, detection of the redshifted
      [C\,\textsc{ii}] $158\mu$m emission line and the continuum {from the Plateau de Bure Interferometer (PdBI)} at
      1.3mm \citep{Venemans2012}, and an upper limit {from the Very Large Array (VLA)} in the radio at
      $1-2$GHz \citep{Momjian2014}. In \S\ref{sec:obs} we present new photometric
      observations with Subaru, \emph{Spitzer}, the Wide-field
      Infrared Survey Explorer (\emph{WISE}), \emph{Herschel}, and
      SCUBA-2. In \S\ref{sec:analysis} we tabulate all the photometric
      measurements of ULAS J1120+0641, and plot the multiwavelength
      SED. We also present an analysis of the SED aimed in particular
      at understanding the contribution of star formation to the
      far-infrared luminosity, and to estimate the bolometric
      luminosity of the source. In \S\ref{sec:discussion} we discuss
      the measurement of the star formation rate in this source, and
      other $z>6$ quasars, and consider the rate of growth of the
      black hole $\dot{M}_{\rm BH}$, and of the stellar mass of the
      bulge $\dot{M}_{\rm bulge}$, and the development of the $M_{\rm
        BH}/M_{\rm bulge}$ relation. We summarise in \S\ref{sec:summary}.

We have adopted a concordance cosmology throughout with \(H_0\) = 70
km s\(^{-1}\) Mpc\(^{-1}\), \(\Omega_{\rm M}\) = 0.3, and
\(\Omega_{\Lambda}\) = 0.7, leading to a luminosity distance for ULAS
J1120+0641 of \(d_L\) = 70.0 Gpc.

\section{New observations and data reduction}
\label{sec:obs}

\begin{table*}[t]
\centering
\caption{New observations of ULAS J1120+0641.}
\label{tab:obs}
\small
\begin{tabular}{cccSrrl}
\hline \hline
Facility & Instrument & Bands & {Wavelength} / \(\mu\)m & \multicolumn{1}{c}{UT date(s) of observation} & Integration time/s & \multicolumn{1}{c}{Program ID}\\ [0.5ex]
\hline
Subaru  & Suprime-Cam & $i^\prime$ & 0.75 & 2013/01/9-11  & 9000 & S12A-010 \\
        &             & $z^\prime$ & 0.89 &               & 4140 &          \\ \hline
UKIRT   & WFCAM       & $H$ & 1.63 & 2011/01/24,26 & 1000 & U/10A/8  \\
        &             & $K$ & 2.20 &               & 1000 &          \\ \hline
\emph{Spitzer} & IRAC        & Ch1 & 3.6  & 2011/07/16    & 2717 & 80114    \\
        &             & Ch2 & 4.5  &               & 2717 &          \\ \hline
\emph{WISE}    &             & W3  &  12  &     2010/06/02-06          & 1170      & ALLWISE  \\
        &             & W4  &  22  &               &   1170    &          \\ \hline
\emph{Herschel}& PACS        &     & 100  & 2012/11/20    & 2592 & 1342255577/8\\
        &             &     & 160  &               & 2592 & \\ \hline
\emph{Herschel}& SPIRE       &     & 250  & 2012/12/09    & 336  & 1342256856 \\
        &             &     & 350  &               & 336  & \\
        &             &     & 500  &               & 336  & \\ \hline
JCMT    & SCUBA-2     &     & 450  & 2012/01/28 -- 03/14 & 31760 & M11BGT01\\
        &             &     & 850  &               & 31760 & \\ \hline
\hline
\end{tabular}
\end{table*}

The new photometric observations of ULAS J1120+0641 are summarised in
Table~\ref{tab:obs}. In the following sub-sections we outline the data
reduction steps and how the photometry was performed. In most cases
the photometric errors are dominated by sky noise, including photon
(Poisson) noise and, at the longest wavelengths, confusion noise. At
all wavelengths the sky noise was estimated by placing apertures on
the sky and measuring the standard deviation in the histogram of
sky-subtracted aperture fluxes, established from measurement of the
negative wing of the Gaussian distribution. Where necessary, any
gradients in the sky were removed before this step. In the case of
high signal-to-noise ratio (S/N) detections, photon noise from the
source was added in quadrature. For the \emph{Herschel} observations
we followed very similar procedures to those used by
\citet{Leipski2013}. All resulting photometry is presented in
Table~\ref{tab:fulldata1}.

At all wavelengths longer than \emph{Spitzer} Ch2 (i.e. beyond
$5\mu$m), the measured flux is less than $2\sigma$. We have recorded
the measured flux, even if negative, and the uncertainty, rather than
quote upper limits, which contain less information. This is useful
when we fit models (\S\ref{sec:analysis}), where the only free parameter is
the normalisation. By fitting to the fluxes all the measurements are
used simultaneously, and combine to constrain the normalisation.

\subsection{Subaru}
\label{sec:subaru}
Images of the field of ULAS J1120+0641 were taken with the Suprime-Cam
instrument \citep{Miyazaki2002} on the Subaru Telescope, in the Sloan
Digital Sky Survey (SDSS) $i^\prime$ and $z^\prime$ filters over 2011
January 9--11. Total exposure times of usable data of 150 min. in
$i^\prime$ and 69 min. in $z^\prime$ were obtained, made up from
individual exposures of, respectively, 180\,s and 300\,s. The data
were reduced and combined using Version 2.0 of the SDFRED package
\citep{Ouchi2004}, and the photometric calibration was applied using
aperture measurements of unsaturated stars in SDSS \citep{Ahn2014}.  A
section of the $z^\prime$ image is reproduced in Fig. \ref{stamp},
left.

\subsection{UKIRT}
\label{sec:ukirt}
The integration time of the original (discovery) UKIRT Infrared Deep
Sky Survey (UKIDSS) Large Area Survey (LAS) $YJHK$ images was 40s
\citep{Lawrence2007}. Deeper $YJ$ photometry of the source was
provided in \citet{Mortlock2011}. We also obtained 500s exposures in
both $H$ and $K$, on both 2011 January 24 and 26. The S/N on the later
date was a factor two better than on the earlier date. These data were
processed by the standard Wide-Field Camera (WFCAM) pipeline. The
frames were calibrated using LAS photometry of bright stars in the
field, and the data from the two nights were combined using
inverse-variance weights. The measurements on the two nights were
consistent with each other. The combined result of $H=18.88\pm0.05$
(Table~\ref{tab:fulldata1}) is not in agreement with the LAS survey
measurement $H=18.24\pm0.14$ (from May 2008), whereas there is good
agreement in $Y$, $J$, and $K$ between both epochs. We were unable to
identify any issues with the data that could explain this
discrepancy. Although there is evidence that the source is variable
\citep{Page2014, Simpson2014}, variability is not the explanation
here, because the $H$ and $K$ observations were taken almost
simultaneously both in the survey and in the follow up, yet the
measured colour has changed significantly
$\Delta(H-K)=0.59\pm0.23$. The newer $H$ value is also discrepant when
compared against a model fit to all the UKIRT data (\S\ref{sec:ad})
and should therefore be considered uncertain.

\subsection{\emph{Spitzer}}
\label{sec:spitzer}
We obtained mid-IR observations using the \emph{Spitzer} InfraRed
Array Camera \citep[IRAC;][]{Fazio2004} in Ch1 ($3.6\mu$m) and Ch2
($4.5\mu$m) in July 2011. By this time \emph{Spitzer} was executing
the Warm Mission, so the longer wavelength channels, Ch3, Ch4, and
MIPS were unavailable. Standard pipeline mosaics were downloaded from
the \emph{Spitzer} Heritage
Archive\footnote{\url{http://sha.ipac.caltech.edu/applications/Spitzer/SHA}}.
A section of the Ch2 image is reproduced in Fig. \ref{stamp},
centre. A median filter was applied to remove a visible gradient in
the sky background in both images. There is a bright star $15\arcsec$
to the NW of the source, marked S in Fig.~\ref{stamp}. Consequently,
a small aperture, of radius $4\arcsec$, was used for the aperture
photometry. An aperture correction was applied to account for this
small size as prescribed by the IRAC Instrument
Handbook\footnote{\url{http://irsa.ipac.caltech.edu/data/SPITZER/docs/irac/iracinstrumenthandbook/}}.

\subsection{\emph{WISE}}
\label{sec:wise}
\emph{WISE} photometry of ULAS J1120+0641 from the All-Sky
mid-infrared survey has been published by \citet{Blain2013}. The W1
($3.4\mu$m) and W2 ($4.6\mu$m) measurements listed there are
superseded by our much deeper \emph{Spitzer} measurements, described
above. \citet{Blain2013} provide only upper limits for the W3
($12\mu$m) and W4 ($22\mu$m) bands. These are quoted as the measured
flux plus two times the sky noise. Since we need fluxes, rather than
limits, we downloaded the ALLWISE W3 and W4 images. These images are
the same as those from the All-Sky mid-infrared survey, but with
improved astrometry. As with the \emph{Spitzer} images a gradient in
the background was removed before estimating the sky noise. The nearby
star that is visible in \emph{Spitzer} Ch1 and Ch2 is very faint at
these wavelengths, so standard \emph{WISE} apertures of radii
$8\farcs25$ (W3) and $16\farcs5$ (W4) were used. Calibration
followed the procedures described in the ALLWISE Explanatory
Supplement\footnote{\url{http://wise2.ipac.caltech.edu/docs/release/allwise/expsup/}}.
Our $2\sigma$ upper limits are consistent with the upper limits quoted
by \citet{Blain2013}.

\subsection{\emph{Herschel}}

\subsubsection{PACS}
\label{sec:pacs}
We observed the source with the Photodetector Array Camera and
Spectrometer \citep[PACS;][]{Poglitsch2010} at 100 and 160
\(\mu\)m. We obtained two maps with scan angles of 70\(\degr\) and 110\(\degr\), with data
taken in miniscanmap mode using a scan speed of $20\arcsec$\,s$^{-1}$ and a scan leg length of 4\(\arcmin\).

\begin{table*}[t]

\small
\begin{adjustbox}{center}
\begin{tabular}{cSccScc}
\hline \hline
Band & {\(\lambda_{\rm obs}\)} & Photometry & \(f_{\nu}\) & {\(\lambda\)} & \(\lambda L_{\lambda}\) & Reference \\ [0.5ex]
     & {\(\mu\)m}  & \(f\)/mag./\(f_{\nu_{\rm obs}}\) &  mJy       &     {\(\mu\)m}   &  \(10^{46}\)erg\,s\(^{-1}\)  \\
\hline \\[-2ex]
X-ray $5-10$ keV & {\(1.8\times10^{-4}\)} & \(<4.2 \times 10^{-16}\) erg/s/cm\(^{2}\) & \(<3.5 \times 10^{-8}\) & {\(2.2\times10^{-5}\)} & < 0.034  & 1 \\

X-ray $2-5$ keV & {\(3.9\times10^{-4}\)} & \(<4.1 \times 10^{-16}\) erg/s/cm$^{2} $ & \(<5.8 \times 10^{-8}\) & {\(4.9\times10^{-5}\)} & < 0.026 & 1 \\

X-ray $0.5-2.0$ keV & {\(1.3\times10^{-3}\)} & \((5.7 \pm 1.2) \times 10^{-16}\) erg/s/cm$^{2}$ & \((1.7 \pm 0.4) \times 10^{-7}\) & {\(1.6\times10^{-4}\)} &\(0.023 \pm 0.005\) & 1 \\

X-ray $0.2-0.5$ keV & {\(3.9\times10^{-3}\)} & \((6.2 \pm 1.7) \times 10^{-16}\) erg/s/cm$^{2}$ & \((8.8 \pm 2.4) \times 10^{-7}\) & {\(4.9\times10^{-4}\)} & \(0.040 \pm 0.011\)& 1 \\

$i'$ & 0.75 & \(28.70\) (AB; SNR = 0.5) & \((1.2 \pm 2.4) \times 10^{-5}\) & 0.092 & \(0.003 \pm 0.006\)& 2 \\

$z'$ & 0.89 & \(23.19 \pm 0.06\) (AB) & \((1.92 \pm 0.11) \times 10^{-3}\) & 0.11 & \(0.38 \pm 0.02\) & 2 \\

$Y$ & 1.03 & \(19.63 \pm 0.04\) (Vega) & \((2.85 \pm 0.11) \times 10^{-2}\) & 0.13 &\(4.86 \pm 0.19\) & 3 \\

$J$ & 1.25 & \(19.22 \pm 0.07\) (Vega) & \((3.14 \pm 0.20) \times 10^{-2}\) & 0.15 &\(4.42 \pm 0.28\) & 3 \\

$H$ & 1.63 & \(18.88 \pm 0.05\) (Vega) & \((2.86 \pm 0.13) \times 10^{-2}\) & 0.20 & \(3.08 \pm 0.14\)& 2 \\

$K$ & 2.20 & \(17.76 \pm 0.04\) (Vega) & \((4.97 \pm 0.18) \times 10^{-2}\) & 0.27 & \(3.97 \pm 0.14\)& 2 \\

\emph{Spitzer} Ch1 & 3.6 & \((63.5 \pm 1.8)\mu\)Jy & \((6.35 \pm 0.18) \times 10^{-2}\) & 0.45 &\(3.10 \pm 0.09\) & 2 \\
\emph{Spitzer} Ch2 & 4.5 & \((58.0 \pm 1.8)\mu\)Jy &  \((5.80 \pm 0.18) \times 10^{-2}\) & 0.56 & \(2.27 \pm 0.07\)& 2 \\
ALLWISE Ch3 & 12 & \((93 \pm 140)\mu\)Jy & \((0.93 \pm 1.40) \times 10^{-1}\) & 1.48 & \(1.4 \pm 2.1\) & 2 \\
ALLWISE Ch4 & 22 & \((0.57 \pm 1.02)\)mJy & \(0.57 \pm 1.02\) & 2.72 & \(4.6 \pm 8.2\) & 2 \\
PACS green & 100 & \((2.5 \pm 1.4)\)mJy & \(2.5 \pm 1.4 \) & 12.4 &\(4.4 \pm 2.5\) & 2 \\
PACS red & 160 & \((4.1 \pm 2.4)\)mJy & \(4.1 \pm 2.4 \) & 19.8 & \(4.5 \pm 2.6\)& 2 \\
SPIRE & 250 & \((-1.2 \pm 7.1)\)mJy & \(-1.2 \pm 7.1 \) & 30.9 & \(-0.8 \pm 5.0\) & 2 \\
SPIRE & 350 & \((4.2 \pm 10.0)\)mJy & \(4.2 \pm 10.0 \) & 43.3 & \(2.1 \pm 5.0\) & 2 \\
SCUBA-2 & 450 & \((-3.9 \pm 4.2)\)mJy & \(-3.9 \pm 4.2\) & 55.7 & \(-1.5 \pm 1.6\) & 2 \\
SPIRE & 500 & \((6.3 \pm 9.6)\)mJy & \(6.3 \pm 9.6 \) & 61.8 & \(2.2 \pm 3.4\) & 2 \\
SCUBA-2 & 850 & \((1.02 \pm 0.92)\)mJy & \(1.02 \pm 0.92 \) & 105 & \(0.21 \pm 0.19 \) & 2 \\
PdBI 235 GHz & 1276 & \(0.61 \pm 0.16 \)mJy & \(0.61 \pm 0.16 \) & 158 &\(0.084 \pm 0.022\) & 4 \\
Radio $1-2$ GHz & 212000 & <23.1$\mu$Jy & \(<2.31 \times 10^{-2}\) & 26200 & \(<1.9\times10^{-5}\) & 5 \\
\hline
\end{tabular}
\end{adjustbox}
\caption{The full SED of ULAS J1120+0641. Column 1 lists the band,
  column 2 the observed wavelength, and column 3 provides the
  photometric measurement in terms of the observational quantity
  usually employed for the particular wavelength regime {\em vis.}
  flux/magnitude/flux density. These are re-expressed in column 4
  uniformly in terms of flux density, $f_{\nu}$. Column 5 lists
  restframe wavelength $\lambda$, and column 6 provides the restframe
  quantity $\lambda L_\lambda$, for the adopted cosmology, where
  $L_\lambda$ is the luminosity density. This quantity was computed
  from the formula $\lambda L_{\lambda}=\nu L_{\nu}=4\pi cf_{\nu}
  d_L^2/\lambda_{\rm obs}$ where $d_L$ is the luminosity distance, and
  $\lambda_{\rm obs}$ is observed wavelength. Column 7 lists the
  source of the photometry. Upper limits (from the literature)
  correspond to 3$\sigma$. See \citet{Page2014} and
  \citet{Simpson2014} for, respectively, \emph{Chandra} and \emph{HST} (F814W,
  F105W, F125W) data that overlap with data listed in this table.
  References: (1) \citet{Page2014}; (2) This Work; (3)
  \citet{Mortlock2011}; (4)\citet{Venemans2012}; (5)
  \citet{Momjian2014}. }
\label{tab:fulldata1}
\end{table*}

\begin{figure*}[t]
\centering
\advance\leftskip-3cm
\advance\rightskip-3cm
\includegraphics[scale = 0.75, trim = 0cm 0.3cm 0cm 0.2cm, clip]{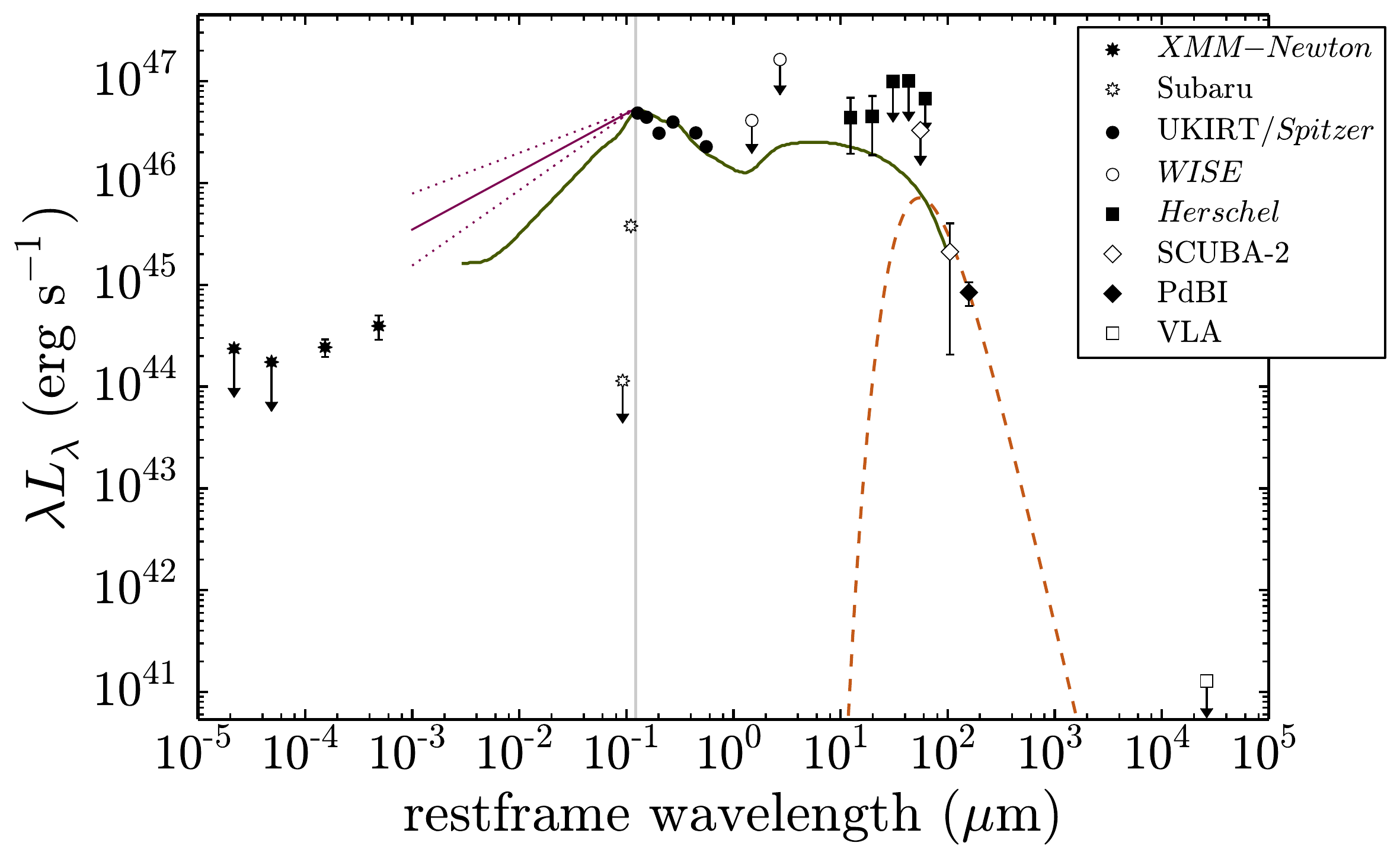}
\caption{Full SED of ULAS J1120+0641. All measurements with
  significance $>1\sigma$ are plotted with error bars. Measurements
  below $1\sigma$ significance are plotted as upper limits (downwards
  arrows) at a value equal to $2\sigma$ (NB not measured
  flux$+2\sigma$). The mean
  quasar template for all SDSS quasars in \citet{Richards2006}, fitted
  to the UKIRT and \emph{Spitzer} points, is shown by the green
  line. The orange dashed line is a modified black body (\(\beta\) =
  1.6 as used by \citet{Beelen2006} and defined in \S\ref{sec:dtbb}, T
  = 47K) fitted to the 1.3mm data point. The vertical grey line
  indicates the rest frame wavelength of Ly\(\alpha\) emission. The
  solid magenta line in the EUV region $10^{-3}-0.12\mu$m is the
  power-law fit from \citet{Telfer2002} for radio-quiet quasars,
  $\alpha=-1.57\pm0.17$, with the uncertainty indicated by the dotted
  lines. The $i'$, $z'$ Subaru observations are not expected to match the intrinsic SED as the continuum is strongly absorbed at these wavelengths.}
\label{sed}
\end{figure*}

Data reduction was performed using the \emph{Herschel} Interactive
Processing Environment \citep[HIPE;][]{Ott2010}, version 11.0.1. Maps
were produced with a custom pixel scale of $1\arcsec$ using
\emph{Herschel} Level 1 data products and the map-making routine
\emph{Scanamorphos} \citep{Roussel2013}. An independent set of maps
was also produced using \emph{unimap} \citep{Piazzo2012}, as yet
unavailable in HIPE. In both cases, the data were processed separately
for the two scan directions before being mosaicked. To help in
visualising the significance of the PACS measurements of the quasar,
we also created an image by combining the 100 and 160 \(\mu\)m images,
binned to $4\arcsec$ pixels. This is reproduced in Fig. \ref{stamp},
right.

The flux of the quasar was measured in the PACS mosaics using aperture
photometry within HIPE. Results were similar for both reductions, and
so were averaged. The radius of the aperture was kept to $7\arcsec$ to
avoid a neighbouring galaxy, marked G in Fig.~\ref{stamp}, and an
appropriate aperture correction was applied \citep{Poglitsch2010}. To
estimate the sky noise we measured the flux in 450 apertures of radius
$7\arcsec$ positioned randomly on the sky
\citep[e.g.][]{Lutz2011,Leipski2013}. The only condition on the
placement of these apertures was that the central pixel should have an
integration time of at least $80\%$ of that of the quasar, established
from the coverage files. We applied a small correction factor to the
sky noise estimate to account for the lower average integration time
of these measurements, compared to the integration time for the
quasar.

\subsubsection{SPIRE}
\label{sec:spire}
We also observed ULAS J1120+0641 with the Spectral and Photometric
Imaging Receiver \citep[SPIRE;][]{Griffin2010} at 250, 350 and 500
\(\mu\)m for 14 repetitions, in small scan map mode. Standard pipeline
images were downloaded from the \emph{Herschel} Science Data
Archive\footnote{\url{http://herschel.esac.esa.int/Science_Archive.shtml}}. Source
extraction was performed on the maps using the built-in HIPE task
\emph{sourceExtractorSussextractor} \citep{Savage2007}, following
recommendations provided in the SPIRE Observer's
Manual\footnote{\url{http://herschel.esac.esa.int/Docs/SPIRE/html/spire_om.html}}. No
sources were detected within $20\arcsec$ of the quasar position.

Detected sources were subtracted from the SPIRE maps to leave a
residual image. The residual maps were recalibrated in units of
Jy/pixel using values from the SPIRE Data Reduction Guide
(DRG)\footnote{\url{http://herschel.esac.esa.int/Data_Processing.shtml}}. We
used apertures of radius $22, 32, 40\arcsec$ for the $250, 350,
500\mu$m wavelength images respectively. We measured small positional
offsets of the SPIRE images relative to \emph{Spitzer}, of a few
arcsec, and corrected for these before measuring the flux at the
position of the quasar. The uncertainty was then estimated in the same
way as for the PACS maps. Our noise measurements are consistent with
the confusion noise measurements presented by \citet{Nguyen2010}.

\subsection{JCMT}
\label{sec:jcmt}
The source was observed with SCUBA-2 at 450 and 850 \(\mu\)m, at the
James Clerk Maxwell Telescope (JCMT), by the instrument team in
Guaranteed Time. The total integration time in each band was nearly 9
hours, and the images reach deeper than the \emph{Herschel}
observations because of the deeper confusion limit, a result of the
larger telescope aperture. The total observing time was split into 13
separate observations. All observations were carried out in weather
bands 1 and 2 (i.e., an optical depth of \(\tau_{\rm 225GHz} <
0.08)\). The data are of uniformly high quality, and we found no
benefit in using lower weights for band 2 relative to band 1 in
creating the mosaics. The raw data were downloaded from the SCUBA-2
archive\footnote{\url{http://www3.cadc-ccda.hia-iha.nrc-cnrc.gc.ca/jcmt/search/scuba2}}.
The data were reduced using the SMURF package developed by
\citet{Chapin2013} and provided by the STARLINK software
project\footnote{\url{http://starlink.jach.hawaii.edu/starlink}}. The
raw files were processed into 13 individual maps, one per observation,
using the SMURF configuration file
\emph{dimmconfig\_blank\_field.lis}. A standard flux correction factor
was applied to the maps to calibrate them in mJy. They were then
mosaicked using the PICARD
package\footnote{\url{http://www.oracdr.org/oracdr/PICARD}}. A
standard matched filter was applied for the source detection.

Some of the SMURF parameters were adjusted from their default
values. The most important of these is the filtering parameter
\emph{filt\_edge\_largescale} (FEL), for which the default value is
200. We experimented with values of FEL$=$150, 200, 250. The noise was
assessed in the usual way from the variance in apertures placed
randomly in the field, within the region of high coverage. For
FEL$=$250, and larger values, rings appear in the final image, and a
bright spot appears at the centre, giving the impression of a
source. But the measured S/N is not significant, showing that the spot is an artefact that is a consequence of an incorrect choice of the value of
FEL. We found that setting FEL$=$150 produced the best results, as
quantified by the measured S/N of two bright sources in the
field. Therefore we settled on this value of FEL, which produced a
flat image.

\section{Analysis}
\label{sec:analysis}

The full SED of ULAS J1120+0641 is provided in
Table~\ref{tab:fulldata1}, comprising the results from \S\ref{sec:obs}
and previously published results, from the references cited in
\S\ref{sec:intro}. The SED is shown in Fig.~\ref{sed}, plotting the
quantity $\lambda L_\lambda(=\nu L_{\nu})$ against restframe
wavelength, covering the range \(0.2\AA - 2.6\)cm.

We are interested in using the SED to measure the bolometric
luminosity $L_{\rm bol}$ of the source and to estimate the
contributions to $L_{\rm bol}$ from the active galactic nucleus (AGN) and from star formation. Below, we measure the contribution to $L_{\rm bol}$ over
the wavelength range $0.12-10^3\mu$m by fitting physically motivated
models to the data. Although six of the new measurements (the two
\emph{WISE}, the three SPIRE, and the $450\mu$m SCUBA-2 values) are
below $1\sigma$, the $1\sigma$ depths reached are comparable to the
expected flux levels, and so the observations provide useful
constraints on the SED at these wavelengths.  The contributions to
$L_{\rm bol}$ at X-ray and radio wavelengths are much smaller than the
uncertainties of the far-IR contribution and so may be neglected in
this analysis.

Also plotted on Fig.~\ref{sed} is the mean SDSS quasar template SED
from \citet{Richards2006}, normalised to the UKIRT and \emph{Spitzer}
observations. The \citet{Richards2006} template appears to provide a
satisfactory fit to the \emph{Herschel} and SCUBA-2 data, and there is
a suggestion that the AGN could be responsible for a significant
proportion of the restframe 158$\mu$m continuum flux (observed
1.3mm). There are, however, a number of reasons why such a conclusion
would be premature. First, as emphasised by \citet{Richards2006},
there is wide variation between the SEDs of quasars, so normalising
the template to the restframe optical part will not necessarily
provide a good fit to the far-infrared region. Second, the longest
wavelength section of the template SED, beyond $50\mu$m restframe,
falls off steeply, but the slope is not well determined. This part of
the template derives from the older template of \citet{Elvis1994}. To
create the AGN template, the correction for host-galaxy light
(i.e. star formation) at these wavelengths is substantial and
uncertain.

An alternative possibility is that most of the measured flux at
restframe 158$\mu$m is from interstellar dust, heated by star
formation. In order to disentangle the contributions to the SED from
star formation and the AGN itself we fit models to the data, following
a similar procedure to that used by \citet{Leipski2013,Leipski2014},
but with fewer free parameters, considering the low S/N of the
data. We use three components to model the SED over the wavelength
range $0.12-10^3\mu$m: an unobscured
accretion disk and broad-line region (AD); a dusty clumpy torus (DT);
and, representing the results of star formation, a modified black body
(BB) of temperature 47K. We also need to include the contribution to
$L_{\rm bol}$ from the extreme ultra-violet (EUV) region, i.e.,
restframe wavelengths $10^{-3}-0.12\mu$m. We now describe
each of these components in turn.

\subsection{EUV component}
\label{sec:euv}
The spectra of quasars at extreme ultra-violet wavelengths are poorly
known because much of the region is unobservable due to absorption by
gas in the host galaxy and in the intervening intergalactic medium. \citet{Telfer2002} have created composite
spectra over the wavelength range $0.05-0.12\mu$m. For radio-quiet
quasars (as ULAS J1120+0641) they find $\alpha=-1.57\pm0.17$, where
$f_{\nu}\propto \nu^{\alpha}$. As shown in Fig.~\ref{sed}, the
\citeauthor{Telfer2002} result, extrapolated to X-ray wavelengths,
lies substantially above the XMM-Newton X-ray
measurements. Nevertheless, it is the slope near the peak that
primarily determines the integrated luminosity.  Therefore we have
integrated the power-law over the wavelength range $10^{-3}-0.12\mu$m,
to obtain $L_{\rm EUV}=8.8\,\pm\,2.2\times10^{46}$erg\,s$^{-1} = 2.3 \pm
0.6\times10^{13}L_{\odot}$, which is adopted as the contribution to
$L_{\rm bol}$ from wavelengths shortward of
Ly$\alpha$. \citet{Page2014} present evidence that ULAS J1120+0641
faded in X-rays between the time the source was observed by \emph{Chandra},
and by \emph{XMM-Newton}, so it is unclear if the X-ray points plotted
represent the typical state.

 Compared to the \citet{Telfer2002} spectrum, the \citet{Richards2006}
 template falls off more steeply and yields $L_{\rm
   EUV}=4.3\times10^{46}$ erg\,s\(^{-1}\). The difference is likely
 mostly due to absorption by neutral hydrogen which was not corrected
 for by \citeauthor{Richards2006} A more sophisticated treatment
 together with a detailed discussion is presented in
 \citet{Krawczyk2013}.

\subsection{AD component}
\label{sec:ad}

In contrast to the EUV regime, the contribution to $L_{\rm bol}$ from
restframe wavelengths between $0.12-1.0\mu$m is very tightly
constrained by the data. For the AD component we use an updated
version of the models of \citet{Maddox2012} over the wavelength range
$0.12-3\mu$m, shown as the red line in Fig.~\ref{sedclose}, cut off as
a power law at longer wavelengths as $f_\nu\propto \nu^2$
\citep{Honig2010}. The photometry at $H$ is discrepant (\S2.2) and was
not used in the fit. We measure $L_{\rm
  AD}=7.9\pm0.2\times10^{46}$erg\,s$^{-1}=2.1\pm0.1\times10^{13}L_{\odot}$.

\subsection{DT component}
\label{sec:dtbb}

For the DT component we employ the clumpy torus models of
\citet{Honig2010}. These models have several free parameters,
including inclination angle, and variables characterising the dust
distribution (radial and vertical scale heights, size and number of
clouds).  We have restricted ourselves to the 960 models with
inclination angle \(\leq45^\circ\), consistent with a Type 1
quasar. Our fundamental assumption in using these models is that they
provide a reasonable representation of the full range of SEDs of the
dusty torus.

\subsection{BB component}

For the BB component we assume an optically thin modified black body
using a dust emissivity power law index of \(\beta = 1.6\). We adopt a
temperature of \(T = 47\)K \citep{Beelen2006}, which is the average
temperature found by these authors in their fits to the SEDs of six
high-redshift quasars, with a range $40-60$K. Since our data are not
good enough to determine the temperature of the BB fit, we have taken
this average value as representative. At the redshift of ULAS
J1120+0641 the temperature of the CMB is \(T_{\rm CMB} =
22\)K. Compared to the 47K BB component, heating by the CMB makes a
negligible contribution to the flux \citep{daCunha2013}. On the other
hand, the CMB does influence the photometry. The measured
sky-subtracted flux from a blackbody source warmer than the CMB will
be measured too low. For a 47K blackbody at $z=7.084$, the measured
background-subtracted flux at restframe 158$\mu$m should be multiplied
by the factor 1.13 to compensate (eqn 18, \citet{daCunha2013}). At
redshift $z=6$ the factor is 1.06. We have chosen not to apply any
corrections to quoted 158$\mu$m fluxes ($\S4$) because the factors
are relatively small compared to the observational errors, and because
the actual blackbody temperature is unknown, meaning that the
correction is quite uncertain.

\subsection{Modelling the DT and BB components}

\begin{figure}[b]
\centering
\advance\leftskip-3cm
\advance\rightskip-3cm
\includegraphics[scale = 0.4, trim = 0cm 0.3cm 0cm 0.3cm, clip]{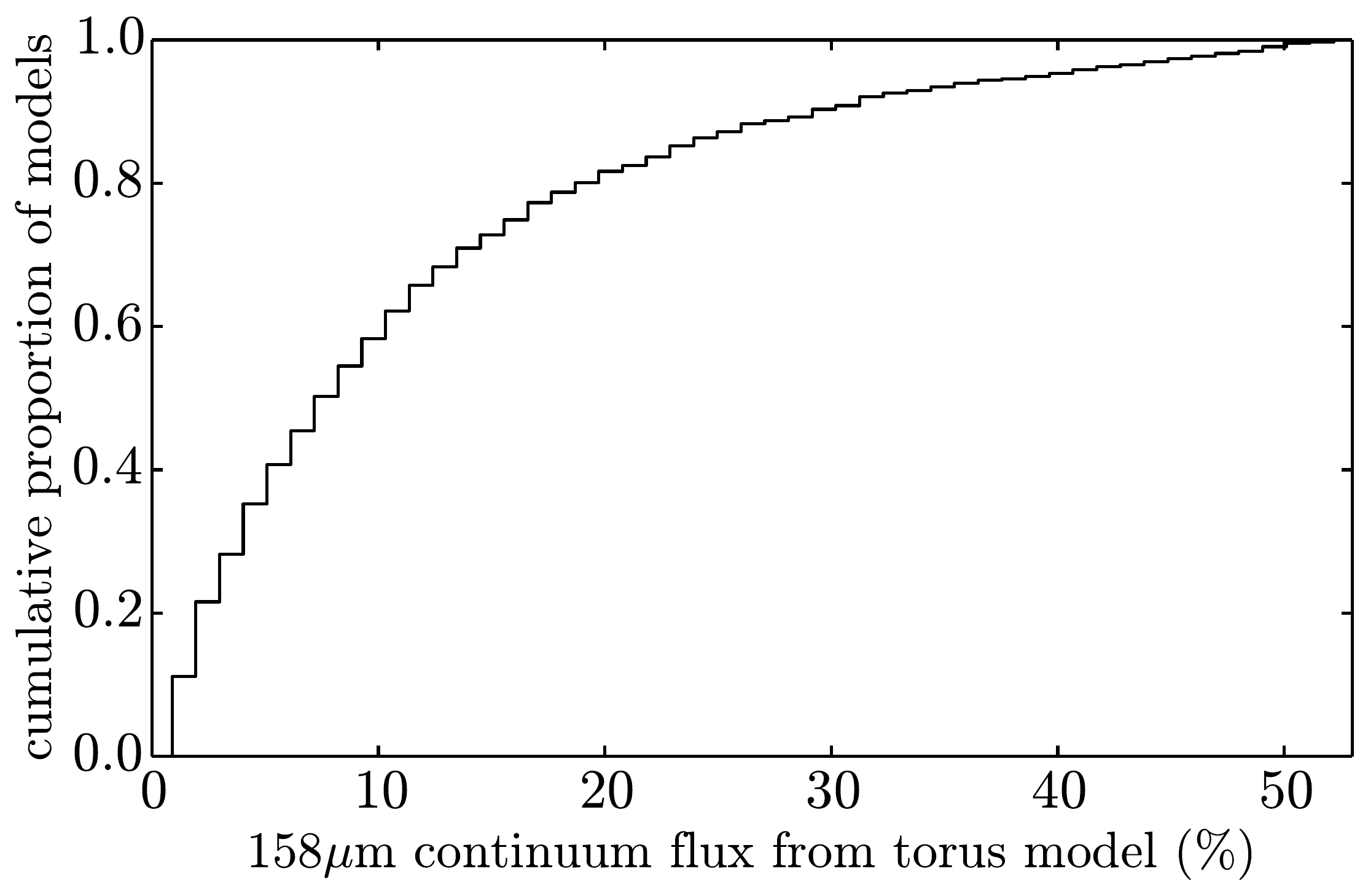}
\caption{The proportion of the 158\(\mu\)m continuum flux that can be
  accounted for by the \citet{Honig2010} torus models.}
\label{hist}
\end{figure} 

\begin{figure*}[t]
\centering
\advance\leftskip-3cm
\advance\rightskip-3cm
\includegraphics[scale = 0.75, trim = 0cm 0.3cm 0cm 0.2cm, clip]{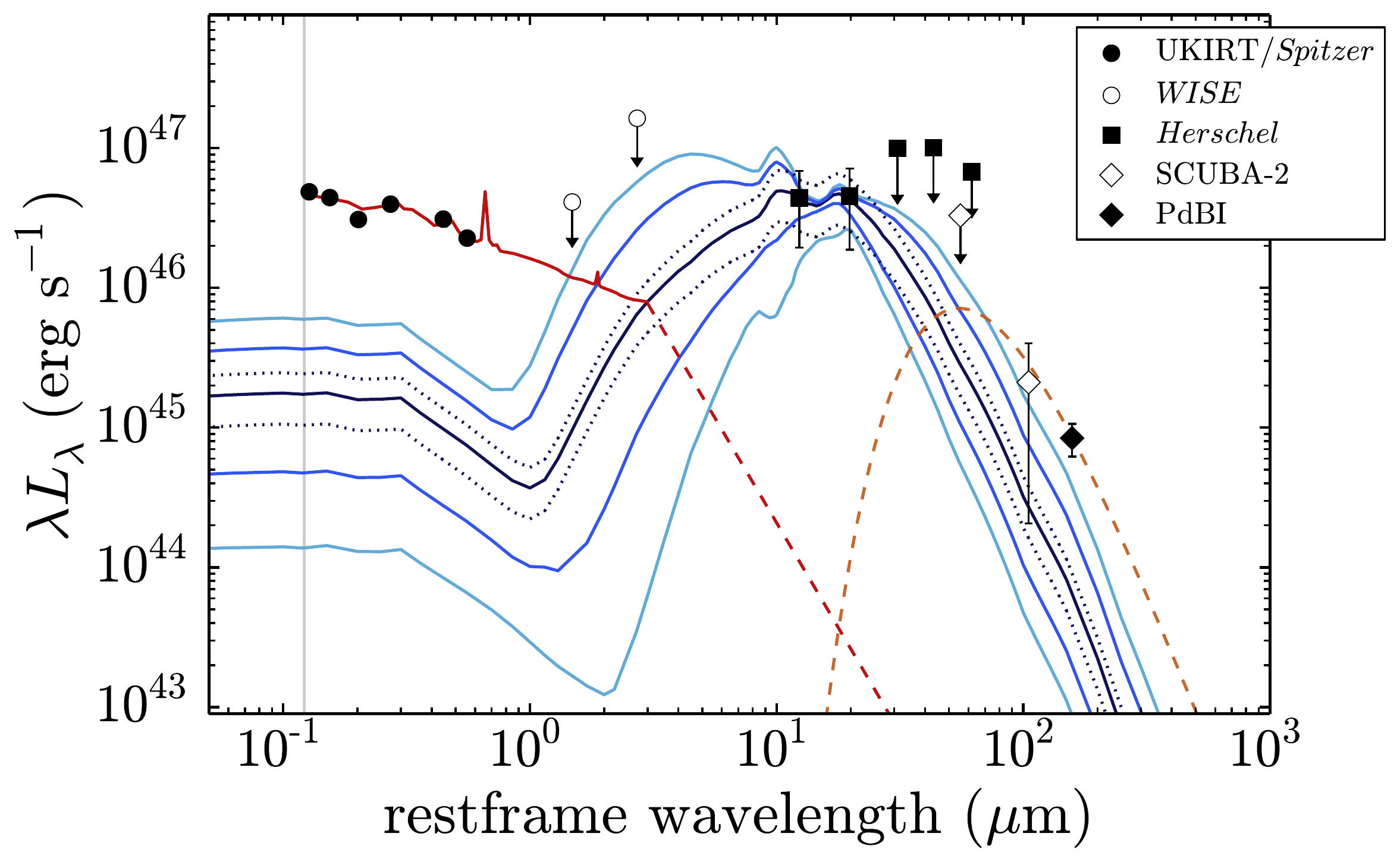}
\caption{Dusty torus model fits. The data points are the same as in
  Fig.~\ref{sed}, with the X-ray and radio ranges cropped. The red
  line represents the AD model, with a break to \(f_{\nu} \propto
  \nu^2\) indicated by the red dashed line beyond 3\(\mu\)m in the
  rest frame \citep{Honig2010}. The AD and BB (orange line) models
  were subtracted from the \emph{WISE}/\emph{Herschel}/SCUBA-2 data,
  and the 960 models of \citet{Honig2010} were fitted to the
  subtracted data. The solid blue lines show the median (darkest) of
  the fits and the $68\%$ and $95\%$ (lightest) ranges. The dotted
  blue lines represent the statistical uncertainty for an individual
  fit.}
\label{sedclose}
\end{figure*}  

To investigate the contribution of the AGN at 158$\mu$m, we begin by
fitting the DT models directly to the \emph{WISE}, \emph{Herschel} and
SCUBA-2 fluxes, with normalisation as a free parameter, by minimising
$\chi^2$ for each model. That is, the best estimate of the
normalisation \(s\) for each torus model is the value that minimises

\begin{equation}
\label{eq:chisq}
\chi^2 = \sum_{i} w_i\,(f_i~-~sM_i)^2~;~i \in \{\rm \emph{WISE}, \emph{Herschel}, SCUBA\text{-} 2\},
\end{equation}
where \(w_i\) is the inverse variance of each observation,
i.e. \(\sigma_i^{-2}\), \(f_i\) is the observed flux, and \(M_i\) is
the value of the model under consideration at the wavelength of the
observation. It is worth reiterating that we fit the models to the
fluxes from \emph{WISE}, PACS, SPIRE and SCUBA-2 rather than limits,
regardless of their significance. Differentiating
Equation~\ref{eq:chisq} with respect to \(s\) and setting the
derivative to zero yields our best estimate of the normalisation for
an individual model:

\begin{equation}
\label{eq:sbest}
s_{\rm best} = \frac{\sum_{i} M_i \, f_i \, w_i}{\sum_{i} M_i^2 \, w_i}
\end{equation} 

The uncertainty in \(s\), \(\sigma_s\), is given by
\begin{equation}
\label{eq:sbestvar}
\sigma_s^{-2} = \sum_{i}M_i^2 \, w_i.
\end{equation}

 For this stage our fitting procedure did not correct for any
 contribution of the AD and BB components over these wavelengths,
 which therefore maximises the DT normalisation, and therefore the
 contribution of the AGN at 158$\mu$m. Since the modified black body
 peaks in $\lambda L_\lambda$ at a restframe wavelength near 55$\mu$m,
 in terms of whether we can rule out a significant contribution by the
 AGN at 158$\mu$m, our approach is conservative. The results are
 provided in Fig.~\ref{hist} where we plot the cumulative distribution
 of the proportion of the 158$\mu$m continuum flux contributed by the
 AGN. Even though the DT normalisation has been maximised, the largest
 contribution at 158$\mu$m of any of the 960 models is $53\%$, while
 $90\%$ of the models contribute $<30\%$. This implies that the BB
 component provides the dominant contribution to the 158$\mu$m
 continuum flux.

\begin{table}[t]
\centering
\caption{The best-fit model SED. Only an extract of the full table which is available in its entirety online is shown here.}
\label{tab:bestfit}
\small
\begin{tabular}{lc}
\hline \hline
{log(restframe wavelength/\(\mu\)m)} &  log(\(\lambda\)L\(_\lambda\)/erg\,s\(^{-1}\))           \\ \hline

-3.00 & 45.542 \\
-2.99 & 45.547 \\
-2.98 & 45.553 \\
-2.97 & 45.558 \\
-2.96 & 45.564 \\

\hline
\end{tabular}
\end{table}

\begin{figure}[h!]
\centering
\advance\leftskip-3cm
\advance\rightskip-3cm
\includegraphics[scale = 0.4, trim = 0cm 0.0cm 0cm 0.cm, clip]{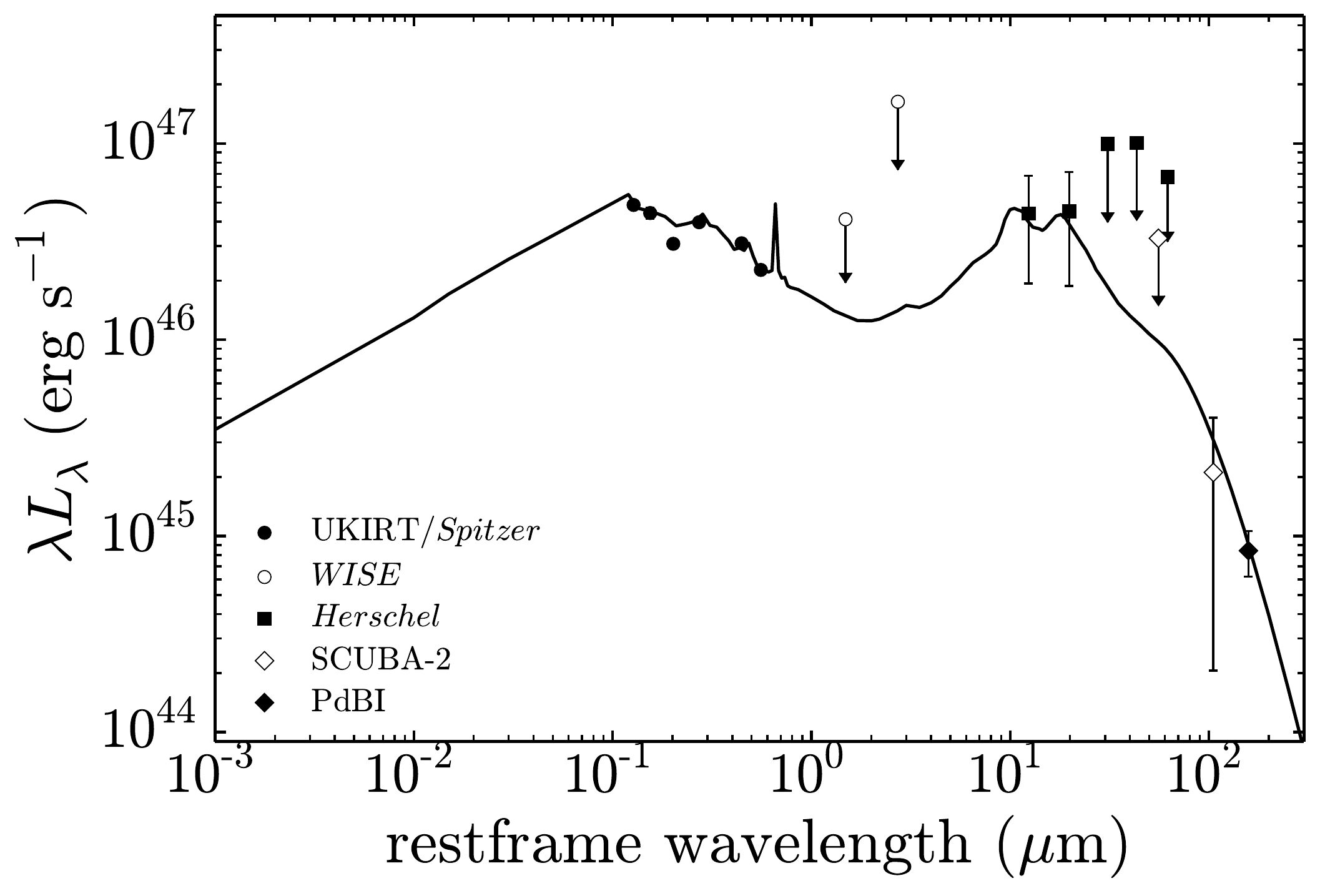}
\caption{The best-fit model SED. Contributions from the EUV, AD, DT
  and BB components are all included. The curve is tabulated in Table~\ref{tab:bestfit}.}
\label{sum}
\end{figure}

Accordingly, we now assume that the contribution
to the SED from star formation is characterised by the BB fit to the
158$\mu$m continuum flux. Further justification for this assumption is
provided in \S\ref{starform}. The BB fit is shown as the dashed curve in
Figs~\ref{sed} and \ref{sedclose}. Integrating under this curve to
determine the BB contribution to $L_{\rm bol}$, we find $L_{\rm
  BB}=7.8\pm2.0\times10^{45}$erg\,s$^{-1}=2.0\pm0.5\times10^{12}L_{\odot}$
($L_{\rm BB}$ is the same as $L_{\rm TIR}$ in \citealt{Venemans2012}).
The BB fit and the AD fit described above were then subtracted from
the observations, and we proceeded by refitting the clumpy torus
models to the subtracted \emph{WISE}/\emph{Herschel}/SCUBA-2 fluxes,
to estimate the DT contribution to the bolometric luminosity.

The results of the DT analysis are illustrated in
Fig.~\ref{sedclose}. The spread in fitted flux at any particular
wavelength, except near the PACS wavelengths, is dominated by the
range of model fits, rather than by the statistical uncertainty in the
normalisation of any fit. For the 960 models, we extract the median
value of $\lambda L_{\lambda}$ as well as the $68\%$ and $95\%$ range
at each wavelength. These are plotted as solid lines in the figure. In
fitting any particular torus model the uncertainty in the
normalisation is typically $40\%$. The dashed lines plotted around the
median fit illustrate this statistical uncertainty. For each model we
integrate over wavelength to obtain $L_{\rm DT}$, and estimate the
uncertainty in this quantity from the range over all models, combined
in quadrature with the statistical uncertainty. The result is $L_{\rm
  DT}=8.4\pm5.3\times10^{46}$\,erg\,s$^{-1}=2.2\pm1.4\times10^{13}L_{\odot}$,
very similar to the measured value of $L_{\rm AD}$ but with a much
larger uncertainty. It is noticeable from Fig.~\ref{sedclose} that the
largest contribution to the uncertainty in $L_{\rm DT}$ comes from the
lack of constraints at restframe wavelengths in the region of $5\mu$m,
where \emph{Spitzer} $24\mu$m observations proved valuable for the
sample of \citet{Leipski2014}. Future observations with the James Webb
Space Telescope will enable us to constrain $L_{\rm DT}$ more
tightly than the current \emph{WISE} observations, with mid-infrared
sensitivity over the \(5-28 \mu\)m wavelength range
\citep{Gardner2006}.

In Fig.~\ref{sum} we show the best-fit model SED, summing the four
components, and the SED is given in Table~\ref{tab:bestfit}.

\subsection{Bolometric luminosity}
\label{lbol}

\begin{table}[t]
\centering
\caption{Components contributing to the bolometric luminosity}
\label{tab:lbol}
\small
\begin{tabular}{lccS}
\hline \hline
component & $L$          &   $L$ & $\%L_{\rm bol}$ \\
          & erg\,s$^{-1}$ &      $L_{\odot}$        & \\ \hline
 EUV      & $8.8\pm2.2\times10^{46}$ & $2.3\pm0.6\times10^{13}$ & \(34 \pm 8\)  \\
 AD       & $7.9\pm0.2\times10^{46}$ & $2.1\pm0.1\times10^{13}$ & \(31 \pm 1\)  \\
 DT       & $8.4\pm5.3\times10^{46}$ & $2.2\pm1.4\times10^{13}$ & \(32 \pm 20\)  \\
 BB       & $7.8\pm2.0\times10^{45}$ & $2.0\pm0.5\times10^{12}$ & {  \(3 \pm 1\) } \\
 \hline \\[-2.25ex]
 $L_{\rm bol}$  & $2.6\pm0.6\times10^{47}$ & $6.7\pm1.6\times10^{13}$    \\
\hline
\end{tabular}
\end{table}

Summing the contributions from the four components above, we obtain
$L_{\rm bol}=L_{\rm EUV}+L_{\rm AD}+L_{\rm DT}+L_{\rm
  BB}=2.6\,\pm\,0.6\times10^{47}$\,erg\,s$^{-1}=6.7\pm1.6\times10^{13}L_{\odot}$.
We tabulate the different contributions to $L_{\rm bol}$ in
Table~\ref{tab:lbol}, expressed in erg\,s$^{-1}$, or $L_{\odot}$, and
as the percentage contribution to $L_{\rm bol}$. 

We now compare the measured value of $L_{\rm bol}$ to the Eddington
luminosity of the black hole. The bolometric luminosity is the total
rate of flow of energy over a sphere centred on the source. The
quantity $L_{\rm bol}$, i.e. the integral under the SED, will equate
to the (true) bolometric luminosity only if the emission is isotropic,
which is almost certainly untrue. This point is discussed by
\citet{Marconi2004} and \citet{Richards2006}. \citet{Marconi2004}
advocate ignoring the far-infrared contribution to $L_{\rm bol}$,
since this energy is effectively counted twice. Referring to
Table~\ref{tab:lbol} we see that the EUV and AD components account for
$0.65L_{\rm bol}$. If these two components suffer from extinction,
however, this sum will underestimate the true bolometric
luminosity. This means that the true bolometric luminosity lies within
the range $0.65-1.0L_{\rm bol}$. For the remainder of the paper we simply
assume that the measured value of $L_{\rm bol}$ is close to the true value.

{The black hole mass is $\sim 2 \times 10^9 M_\odot$ \citep{Mortlock2011,deRosa2014}. While variations in the data and analysis methods provide different formal estimates, in both cases the uncertainty is dominated by the 0.55 dex uncertainty in the normalisation of the \citet{Vestergaard2009} scaling relation. The Eddington ratio is hence $L / L_{\rm Edd} \simeq 1$, with a comparably large relative uncertainty.}

\section{Discussion}
\label{sec:discussion}

\subsection{Star formation}
\label{starform}

Modelling the SED of ULAS J1120+0641 using four components shows the
BB component makes only a small contribution to $L_{\rm bol}$, and
therefore our estimate of $L_{\rm bol}$ is not sensitive to our
assumption that the BB fit to the $158\mu$m continuum point is a good
representation of the contribution of star formation to the
SED. Nevertheless the issue of star formation in high-redshift quasars
bears on the questions of the timescale for reaching supersolar
metallicities in the broad line region, and of the origin of the
correlation between black hole mass and host-galaxy bulge
mass. Therefore we now consider this issue further.

\begin{table*}[t]
\centering
\caption{Measurements of equivalent width of [C\textsc{ii}] for $z>5.7$ quasars}
\label{tab:ewcii}
\small
\begin{tabular}{llllllc}
\hline \hline
Source & Redshift & \multicolumn{1}{c}{flux [C\textsc{ii}]} &
\multicolumn{1}{c}{$f_\nu(158)$} & \multicolumn{1}{c}{EW
  [C\textsc{ii}]} & $\lambda L_{\lambda}(158)$ & Ref. \\ [0.5ex] 
       &          & \multicolumn{1}{c}{Jy\,km\,s$^{-1}$}   &
\multicolumn{1}{c}{mJy} & \multicolumn{1}{c}{$\mu$m} & $10^{45}$
erg\,s$^{-1}$ &    \\ \hline 
\hline
J0129-0035 & 5.779 & $1.99 \pm0.12 $ & $2.57\pm0.06 $ & $0.41\pm0.03$ & 2.6  & 3 \\
J0210-0456 & 6.432 & $0.269\pm0.037$ & $0.12\pm0.035$ & $1.18\pm0.38$ & 0.14 & 4 \\
J1044-0125 & 5.785 & $1.70 \pm0.30 $ & $3.12\pm0.09 $ & $0.29\pm0.05$ & 3.2  & 3 \\
J1120+0641 & 7.084 & $1.03 \pm0.14 $ & $0.61\pm0.16 $ & $0.89\pm0.26$ & 0.84 & 1 \\
J1148+5251 & 6.419 & $3.9  \pm0.3  $ & $4.5 \pm0.62 $ & $0.46\pm0.07$ & 5.4  & 2 \\
J1319+0950 & 6.133 & $4.34 \pm0.60 $ & $5.23\pm0.10 $ & $0.44\pm0.06$ & 5.8  & 3 \\
J2054-0005 & 6.039 & $3.37 \pm0.12 $ & $2.98\pm0.05 $ & $0.60\pm0.02$ & 3.3  & 3 \\
J2310+1855 & 6.003 & $8.83 \pm0.44 $ & $8.91\pm0.08 $ & $0.52\pm0.03$ & 9.6  & 3 \\
\hline
\end{tabular}
\caption*{References: (1) \citet{Venemans2012}; (2)
\citet{Walter2009}; (3) \citet{Wang2013}; (4) \citet{Willott2013}.}
\end{table*}

{In starburst galaxies most of the energy from young stars is absorbed
by dust and re-emitted at far-infrared wavelengths. The best estimate
for the star formation rate (SFR) in starburst galaxies uses the relation
SFR$(M_\odot \,{\rm yr^{-1}}) = 4.5\times 10^{-44} L_{\rm TIR} ({\rm
  erg\,s}^{-1}),$ from \citet{Kennicutt1998}, where $L_{\rm TIR}$
refers to the luminosity integrated over the IR
spectrum between $8-1000\,\mu$m. A far-infrared bump is ubiquitous in the SEDs
of quasars, but the relative contributions from star formation or the
AGN, and the luminosity dependence of this ratio is a matter of
extensive debate, with no clear consensus (e.g. \citet{Haas2003,
  Netzer2007, Wang2011}). Consequently, $L_{\rm TIR}(8-1000\,\mu$m)
only provides an upper limit to the SFR in quasars. By integrating the
SED fits from the previous section over the wavelength range
$8-1000\,\mu$m and applying the Kennicutt relation we obtain the
distribution for the upper limit on star formation for ULAS J1120+0641, which is found to be $2700\pm400\,M_\odot$yr$^{-1}$.}

{\citet{Sargsyan2012} (hereafter S12) and \citet{Sargsyan2014}
  (hereafter S14) have investigated the use of the [C\,\textsc{ii}]
  $158\,\mu$m line luminosity as a measure of star formation rate. From
  PACS spectroscopy of 130 low-redshift galaxies, including AGN,
  starbursts, and composites, they find that the [C\,\textsc{ii}] line
  strength is closely correlated with the strength of certain mid-IR
  line features that are indicators of star formation, particularly
  the PAH 11.3\,$\mu$m feature and the [Ne\,\textsc{ii}]
  line. Furthermore, the line ratios are independent of classification
  into starburst or AGN. S14 conclude that the relation \(\log\,[{\rm SFR}/M_\sun\,{\rm yr}^{-1}] = \log [L([{\rm C}\,\textsc{ii}])/L_\sun]-7.0\pm0.2\)
  measures the SFR in an
  individual source. In addition S12 argue
  that the far-infrared luminosity $L_{\rm TIR}$ is only a reliable
  SFR indicator in the absence of an AGN, and that the contribution
  from the AGN explains why the luminosity ratio $L_{\rm
    [C\,\textsc{ii}]}/L_{\rm TIR}$ falls at high luminosities, $L_{\rm
    TIR}>10^{12}L_{\odot}$. In contrast, S14 find that the restframe
  $158\,\mu$m continuum luminosity $L_{158}$ is correlated with the
  [C\,\textsc{ii}] luminosity, for AGN as well as starbursts. Therefore
  they argue that $L_{158}$ is also a reliable SFR indicator given by the relation \(\log\,[{\rm SFR}/M_\sun\,{\rm yr}^{-1}] = \log [\lambda L_{\lambda}(158\mu{\rm m})/{\rm erg\,s}^{-1}]-42.8\pm0.2\).}

{The [C\,\textsc{ii}] and $158\,\mu$m continuum SFR relations proposed by S14
  are potentially important, because they apply to AGN as well as
  starbursts. These SFR relations are based on an analysis of
  low-redshift sources, nearly all at $z<0.1$, and it is unclear if
  they hold for the high redshift and high luminosity of ULAS
  J1120+0641.  Nevertheless if both $L_{\rm[C\,\textsc{ii}]}$ and
  $L_{158}$ are reliable SFR indicators for luminous sources, both AGN
  and starbursts, over a range of luminosities and redshifts, a
  prediction is that the ratio $L_{\rm [C\,\textsc{ii}]}/L_{158}$,
  i.e. the [C\,\textsc{ii}] equivalent width (EW), should be similar
  for quasars at very high redshift $z>6$.}

{S14 measure a median (restframe) [C\,\textsc{ii}] EW of $1.0\,\mu$m
  for low-redshift starbursts. In Table \ref{tab:ewcii} we collect
  measurements from the literature of the [C\,\textsc{ii}] EW for
  eight $z>5.7$ quasars. The unweighted mean EW for the eight quasars
  is $0.6\mu$m, and the inverse-variance-weighted mean is
  $0.5\mu$m. This is only a factor two smaller than the value of S14,
  despite the much higher $158\,\mu$m luminosities of some of the
  sources, and the large redshift difference. The suggestion is that the
  [C\,\textsc{ii}] line luminosity and restframe $158\mu$m continuum
  luminosity are promising tools for estimating the SFR in luminous
  high-redshift quasars, and that additional measurements over a range of
  redshifts and luminosities would be useful in the future.}

{ULAS J1120+0641 has a measured [C\,\textsc{ii}] restframe EW of
  $0.89\pm0.26\,\mu$m, consistent with the predicted value of $1.0\,\mu$m,
  and therefore providing support for the BB fit used in the previous
  section. Applying the S14 SFR relations to the measurements for
  ULAS J1120+0641, from \citet{Venemans2012}, and listed in
  Table~\ref{tab:ewcii}, we find SFR$_{\rm [C\,\textsc{ii}]} =
  60-220\,{M}_{\odot}$yr$^{-1}$ and SFR$_{158} =
  60-270\,{M}_{\odot}$\,yr$^{-1}$.}

\subsection{The ratio $M_{\rm BH}/M_{\rm bulge}$ }
\label{magorrian}

We now consider the relationship between the mass of the central black
hole in galaxies, $M_{\rm BH}$, and the stellar mass of the bulge,
$M_{\rm bulge}$. An analysis by \citet{Haring2004} of 30 nearby
galaxies yielded a tight relation between the two quantities. The
relation is almost linear, with a fiducial ratio $M_{\rm BH}/M_{\rm
  bulge}\simeq 1.4\times 10^{-3}$. If we equate the quasar SFR to the
rate of increase of the bulge stellar mass, $\dot{M}_{\rm bulge}$, we
are interested in computing the rate of increase of the black hole
mass, $\dot{M}_{\rm BH}$, and comparing the ratio of the growth rates
to the mass ratio observed in galaxies today. In other words, at the
time observed, is the black hole or the bulge growing faster relative
to the mass ratio measured in the local universe?

The luminosity of a black hole is given by $L_{\rm bol} = \eta \dot{M}
c^2$, where $\eta$ is the black hole efficiency and $\dot{M}$ is the
accretion rate. We consider mass that is not converted to energy
through accretion as contributing to the increase of the black hole
mass, i.e., $\dot{M}_{\rm BH} = \frac{(1 - \eta)}{\eta}\frac{L_{\rm
    bol}}{c^2}$. From our value of $L_{\rm bol} = 2.6 \times
10^{47}\,\rm{erg\,s^{-1}}$, and assuming an efficiency $\eta = 0.1$,
we estimate $\dot{M}_{\rm BH} = 40\,{M}_{\odot}$yr$^{-1}$. Adopting a SFR of $\sim
200{M}_{\odot}\textrm{yr}^{-1}$, we find ${\dot{M}_{\rm BH}}/{\dot{M}_{\rm
    bulge}} \simeq {40\,{M}_{\odot}\rm{yr}^{-1}}/\,{200\,{M}_{\odot}\rm{yr}^{-1}} = 0.2$. Comparing this quantity to the
local mass ratio ${M_{\rm BH}}/{M_{\rm bulge}}\simeq 1.4\times
10^{-3}$, the black hole was
growing in mass more than 100 times faster than the stellar bulge,
relative to the mass ratio measured in the local universe.

\section{Summary}
\label{sec:summary}

Combining published measurements and new observations, we have compiled
a full multi-wavelength SED for the z = 7.1 quasar ULAS J1120+0641,
summarised in Table~\ref{tab:fulldata1}, and plotted in Fig.~\ref{sed}. In particular,
the SED included new observations in the far-infrared and sub-mm.

We now summarise the main results of the paper:

\begin{enumerate}
\item Based on an analysis that used the dusty torus models of
\citet{Honig2010}, we find that the torus does not contribute a
significant fraction of the restframe 158$\mu$m continuum flux, which
we ascribe instead to star formation.
 
\item From the model fits we measure a bolometric luminosity of
$L_{\rm bol}=2.6\pm0.6\times10^{47}$erg\,s$^{-1}=6.7\pm1.6\times10^{13}L_{\odot}$, 
where the main source of uncertainty is the lack of deep observations
at $\sim 40 \mu$m, corresponding to 5$\mu$m in the rest frame of the quasar.

\item {A comparison of the [C\,\textsc{ii}] EWs in a sample of $z>5.7$
  quasars with the measured values for starburst galaxies in the local universe
suggests that the [C\,\textsc{ii}] line luminosity and restframe
$158\,\mu$m continuum luminosity are promising indicators for estimating the
SFR in luminous high-redshift quasars.}

\item Based on the [C\,\textsc{ii}] luminosity and the restframe
  158\(\mu\)m continuum luminosity we estimate a SFR in ULAS J1120+0641 of
  $60-270\,{M}_{\odot}$yr$^{-1}$.

\item We find that, at the time observed, the black hole was growing
  in mass more than 100 times faster than the stellar bulge, relative
  to the mass ratio measured in the local universe.

\end{enumerate}

\begin{acknowledgements}

We are grateful to Cristian Leipski for correspondence on PACS and
SPIRE data reduction and photometry, to Bruno Altieri who produced the
{\em unimap} PACS image, to Ros Hopwood for additional expert
advice on the \emph{Herschel} data, to Jim Geach for advice on processing
the SCUBA-2 data, and to Tom Kerr for much help through the UKIRT service
programme. \\
SW gratefully acknowledges the support of the Leverhulme Trust through
the award of a Leverhulme Research Fellowship. \\
BPV acknowledges funding through the ERC grant "Cosmic Dawn". \\
Based in part on data collected at Subaru Telescope, which is operated by the National Astronomical Observatory of Japan. The United Kingdom Infrared Telescope was operated by the Joint
Astronomy Centre on behalf of the Science and Technology Facilities
Council of the U.K. Some of the data reported here were obtained as
part of the UKIRT Service Programme. This work is based in part on
observations made with the \emph{Spitzer} Space Telescope, which is operated
by the Jet Propulsion Laboratory, California Institute of Technology
under a contract with NASA. This publication makes use of data
products from the Wide-field Infrared Survey Explorer, which is a
joint project of the University of California, Los Angeles, and the
Jet Propulsion Laboratory/California Institute of Technology, funded
by the National Aeronautics and Space Administration. \emph{Herschel} is an
ESA space observatory with science instruments provided by
European-led Principal Investigator consortia and with important
participation from NASA. The James Clerk Maxwell Telescope has historically been operated by the Joint Astronomy Centre on behalf of the Science and Technology Facilities Council of the United Kingdom, the National Research Council of Canada and the Netherlands Organisation for Scientific Research. Additional funds for the construction of SCUBA-2 were provided by the Canada Foundation for Innovation. 

\end{acknowledgements}

\end{document}